\documentclass[apj]{emulateapj}
\usepackage{color}
\usepackage{epsf}
\usepackage{ulem}
\usepackage{graphicx}

\newcommand{\Om}{\Omega_m}
\newcommand{\Ob}{\Omega_b}

\newcommand{\OL}{\Omega_\Lambda}

\newcommand{\Mpc}{{\rm\;Mpc}}
\newcommand{\Gpc}{{\rm\;Gpc}}

\newcommand{\wl}{\omega_{\ell}(r_s)}
\newcommand{\wnot}{\omega_{0}(r_s)}

\begin{document}
\title{A New Statistic for Analyzing Baryon Acoustic Oscillations}
\shorttitle{New Statistic for Analyzing BAO}

\author{
X. Xu\altaffilmark{1},
M. White\altaffilmark{2,3},
N. Padmanabhan\altaffilmark{4},
D.~J. Eisenstein\altaffilmark{1},
J. Eckel\altaffilmark{1},
K. Mehta\altaffilmark{1},
M. Metchnik\altaffilmark{1},
P. Pinto\altaffilmark{1},
H.-J. Seo\altaffilmark{5}}

\begin{abstract}
We introduce a new statistic $\wl$ for measuring and analyzing large-scale
structure and particularly the baryon acoustic oscillations.  $\wl$ is a
band-filtered, configuration space statistic that is easily implemented
and has advantages over the traditional power spectrum and correlation
function estimators. Unlike these estimators, $\wl$ can localize most
of the acoustic information into a single dip at the acoustic scale
while also avoiding sensitivity to the poorly constrained large scale
power (i.e., the integral constraint) through the use of a localized and
compensated filter. It is also sensitive to anisotropic clustering through
pair counting and does not require any binning of data.  We measure the
shift in the acoustic peak due to nonlinear effects using the monopole
$\wnot$ derived from subsampled dark matter catalogues as well as from
mock galaxy catalogues created via halo occupation distribution (HOD)
modeling. All of these are drawn from $44$ realizations of $1024^3$
particle dark matter simulations in a 1$h^{-1}\Gpc$ box at z=1. We compare
these shifts with those obtained from the power spectrum and conclude
that the results agree. We therefore expect that distance measurements
obtained from $\wnot$ and $P(k)$ will be consistent with each other. We
also show that it is possible to extract the same amount of acoustic
information by fitting over a finite range using either $\wnot$ or $P(k)$ 
derived from equal volume surveys.
\end{abstract}


\altaffiltext{1}{Steward Observatory, University of Arizona,
                933 N. Cherry Ave., Tucson, AZ 85121; xxu@as.arizona.edu}
\altaffiltext{2}{Departments of Physics and Astronomy, University of California, Berkeley, CA 94720}
\altaffiltext{3}{Lawrence Berkeley National Laboratory, 1 Cyclotron Road, Berkeley, CA}
\altaffiltext{4}{Department of Physics, Yale University, New Haven, CT 06511}
\altaffiltext{5}{Center for Particle Astrophysics, Fermi National Accelerator 
Laboratory, P.O. Box 500, Batavia, IL 60510-5011}

\section{Introduction}\label{sec:intro}
Baryon acoustic oscillations (BAOs) are relic imprints on the baryon
distribution left behind by the interaction between radiation and matter
in the primordial universe.  The large amount of radiation pressure
due to high temperatures in the early universe pushed matter apart as
it congregated under the influence of gravity. Oscillating sound waves
were set up in the primordial plasma due to these two competing effects
\citep{PY70, SZ70, BE84, H89, HS96, HW96, EH98}.  The distance traveled
by these sound waves before recombination is about 150 comoving Mpc
and is known as the acoustic scale. When the photons stream off, the
baryons are deposited at these characteristic separations and hence we
would expect excesses with this separation in the baryon distribution
today. This characteristic scale makes for a very useful standard ruler
\citep{EH98, EHT98, E03, BG03, HH03, L03, SE03, M04, Aea05}.

As the universe evolves, primordial density perturbations begin to
grow nonlinearly, especially on small scales \citep{JB94, Mea99, MW99,
Sea99}. This causes the acoustic scale to become slightly shifted from
its predicted linear theory position \citep{SE05, Hea07, Ma07, Aea08,
CS08, Sanchez08, Sea08, Smith08, PdW09}. One must calibrate this shift
before the acoustic scale can be used as a high precision standard ruler.

Previous calibrations and other analyses of the acoustic signature have
been conducted using the traditional power spectrum and correlation
function methods. Although analytically they are both perfectly adequate,
the estimators used to derive them from simulations and observational
data are subject to numerous difficulties.

The largest survey scales are always poorly constrained due to effects
such as the integral constraint making it a challenge to estimate the
correlation function $\xi(r)$ accurately at these scales. The integral
constraint arises due to the fact that we do not know the cosmic number
density of any population of mass tracers \citep{deLaea88, BF91, PN91,
Hamil93}. Many techniques used to estimate $\xi(r)$ \citep{Peebles73,
Sharp79, Hewett82, BA88, LS93, Hamil93} take the number density of tracers
in the survey volume to be the true number density. This assumption
effectively ignores all power at scales larger than the survey size while
simultaneously increasing the correlation between scales smaller than
the survey size which causes the off-diagonal covariance matrix terms
to be larger than they would be otherwise.

Limited survey volume and awkward survey boundaries are the major
concerns when trying to estimate the power spectrum $P(k)$. Typically,
the measured power spectrum is a convolution of the window function,
the Fourier transform of the selection function of the survey, and the
underlying true power spectrum \citep{FKP94, Park92, BF91, PN91, KP91}.
Therefore, these $P(k)$ estimators are biased. In the limit of infinite
volume, the window function should be a delta function. However, real
surveys have finite volume and hence the window function has a finite
albeit very small width. This induces an artificial smoothing at small
separations in $k$ when attempting to deconvolve the window function
from the observed density field. Pair counting estimators, like those
for $\xi(r)$, avoid this issue because the relative positions of all
tracer particles and hence the true distribution is recorded.

Estimating $\xi(r)$ and $P(k)$ also requires the binning of data. When any
binning process is invoked, one must carefully define any averaging used
to derive the value in each bin. To reduce these errors, bin sizes are
made smaller which increases the dimensionality of the covariance matrix,
making it even more difficult to estimate. Also, survey boundaries need
to be addressed with special care.

In the study of large scale structure, we are also interested in
any anisotropies in the distribution of objects. These can result
from the bulk motions of objects \citep{Peebles80, DP83, Kaiser87} 
as well as assumed models for the Hubble parameter $H(z)$ and the angular
diameter distance $D_A(z)$ while calculating object separations along the 
line of sight (LOS) and transverse directions respectively. 
Since, at large scales we would expect the distribution to be isotropic, any
anisotropy can be deconstructed into velocity field information which in
turn provides us with constraints on $\Omega_m$, the derivative of the
growth function $f = d\mathrm{ln}D/d\mathrm{ln}a$ and the anisotropic
parameter $\beta = f/b$, where $b$ is the galaxy bias \citep{Kaiser87,
Hamil98, Pea01, Hea03, Tea02, Zea02, SP08, PW09, WSP09}. After modeling
these anisotropies, any residual anisotropies can be used to constrain
$D_A(z)H(z)$ \citep{AP79}. If our assumed $H(z)$ or $D_A(z)$ models
are incorrect then the distribution of objects will still appear
anisotropic after the bulk motion effects are removed.  The magnitude
of this residual anisotropy can be used to infer the true underlying
cosmology \citep{PW08, Oea08}.

Anisotropic information can be extracted from the redshift-space
correlation function and to a lesser extent from the redshift space power
spectrum \citep{Szalay98}. If one imagines a wide angle survey, $P(k)$
estimators that rely on a Fourier transform from a Cartesian grid will
suffer because an arbitrary wave vector ($\vec{k}$) will not necessarily
be parallel to the LOS \citep{CFW94, CFW95, ZH95, Szapudi04}. This means
that each $\vec{k}$ mode will contain information about both the LOS
and transverse distributions. After averaging to obtain the spherically
averaged power spectrum, any anistropies in the distribution of survey
objects will have been erased. Instead of taking a spherical average, one
can also analyze the full 3D $P(\vec{k})$ through Legendre decomposition
into radial and angular components. However, an infinite sum is required
and applicability is limited to the linear regime \citep{HT95, Pea04}.
Pair count estimators for $\xi(r)$, on the other hand, record the
distribution of tracers accurately because they record each pair's angle
to the LOS as well as their separation. This means that any anisotropies
in the distribution of galaxies will become obvious. Statistics sensitive
to anisotropic clustering are desirable as they offer us a means to
probe the underlying cosmology.

It is also aesthetically pleasing, to localize the acoustic information
into a single feature at the acoustic scale.  This is true for $\xi(r)$
but not for $P(k)$, which has oscillatory acoustic features.

Although all of the above mentioned disadvantages of $\xi(r)$ and $P(k)$ are
minor, it is still beneficial to derive a new statistic that does away with 
as many of the above setbacks as possible. $\wl$ is an example of such an
alternative.

The organization of this paper will be as follows. In \S\ref{sec:omega},
we introduce $\wl$, including its properties and computation. In
\S\ref{sec:simmeth}, we describe the simulations, halo occupation models
and analysis methods we use to derive acoustic peak shifts through
implementation of the monopole $\wnot$.  We demonstrate the mutual
consistency between the peak shifts measured from the same simulations
using $\wnot$ and $P(k)$ in \S\ref{sec:comppk}.  This is indicative
of the agreement we expect between distance measures from $\wnot$ and
$P(k)$. We also show that with a reasonable finite fitting range and
our current choice of filter for computing $\wl$, we can extract the
same amount of acoustic information using either $\wnot$ or $P(k)$ from
equal volume surveys.  Finally, we state the main results of this paper
in \S\ref{sec:theend}. Details of the matter and galaxy model $P(k)$
results discussed in this paper can be found in the companion papers
\citet{Sea10} and \citet{Mea09} respectively (hereafter referred to as 
S\&M collectively).

\section{The $\wl$ Statistic}\label{sec:omega}

\subsection{Equations and Properties}\label{sec:om_eq}
We expand the angle dependence of the power spectrum and correlation
function out as a series of Legendre polynomials in 
$\mu = \hat{r}\cdot\hat{z} = \cos(\theta)$, where $\theta$ is the LOS angle:
\begin{eqnarray}
  \xi(r,\mu)&\equiv&
  \sum_\ell \xi_\ell(r)L_\ell(\mu) \\
  \Delta^2(k,\mu) &\equiv& \frac{k^3 P(k,\mu)}{2\pi^2}
  = \sum_\ell \Delta_\ell^2(k) L_\ell(\mu)
\end{eqnarray}
so that
\begin{equation}
  \xi_\ell(r) = i^\ell \int\frac{dk}{k}\ \Delta_\ell^2(k)j_\ell(kr)
\end{equation}
where $j_\ell$ is the spherical Bessel function of order $\ell$ and $L_\ell$ 
is the Legendre polynomial of order $\ell$.

Imagine we have a filter, $W_\ell(r,\mu,r_s)=W_\ell(r,r_s)L_\ell(\mu)$, 
which we take to be compact and compensated ($\int r^2\,dr\ W_\ell(r,r_s)=0$) 
with a characteristic scale $r_s$. 
We define our statistic as the redshift-space correlation function,
$\xi_s(r,\mu)$, convolved with the filter as a function of filtering scale
$r_s$.  
\begin{eqnarray}
  \omega_\ell(r_s) &\equiv&
\label{eqn:omxidef}
  i^\ell \int d^3r\ \xi_s(r,\mu) W_\ell(r,r_s) L_\ell(\mu)\\
  &=& \frac{4\pi i^\ell}{2\ell+1}\int r^2\,dr\ \xi_\ell(r)W_\ell(r,r_s)\\
\label{eqn:omdef}
  &=& \int\frac{dk}{k}\ \Delta_\ell^2(k) \widetilde{W}_\ell(k,r_s)
\end{eqnarray}
with
\begin{equation}
  \widetilde{W}_\ell(k,r_s) \equiv (-1)^\ell \frac{4\pi}{2\ell+1}
  \int r^2\,dr\ W_\ell(r,r_s) j_\ell(kr)
\end{equation}
where the $i^\ell$ has been inserted for later convenience.  By making the
filter compensated, we reduce the sensitivity to the poorly constrained
power at large scales and the dependence on the uncertain mean density in
the sample. The correlation function is defined such that $\xi(r,\mu)+1
\propto n^{-2}$ \citep{Peebles80}. Integrating the left-hand side of
this equation against $W_\ell(r,r_s)L_\ell(\mu)d^3r$ results in $\wl$
scaling directly with $n^{-2}$ following equation (\ref{eqn:omxidef}). The
constant term integrates to $0$ as the filter is compensated. Hence,
any uncertainty in $n$ enters as a pure multiplicative offset in $\wl$,
which is less likely to overwhelm the acoustic signature at large scales.
This in essence, eliminates sensitivity to the integral constraint,
which is a small effect to begin with.  This feature of the filter also
makes the statistic measured in different subvolumes of a survey more
independent. We expect that this will make internal error estimates from
methods such as bootstrap or jackknife more robust \citep{Pea09}.

\begin{figure}
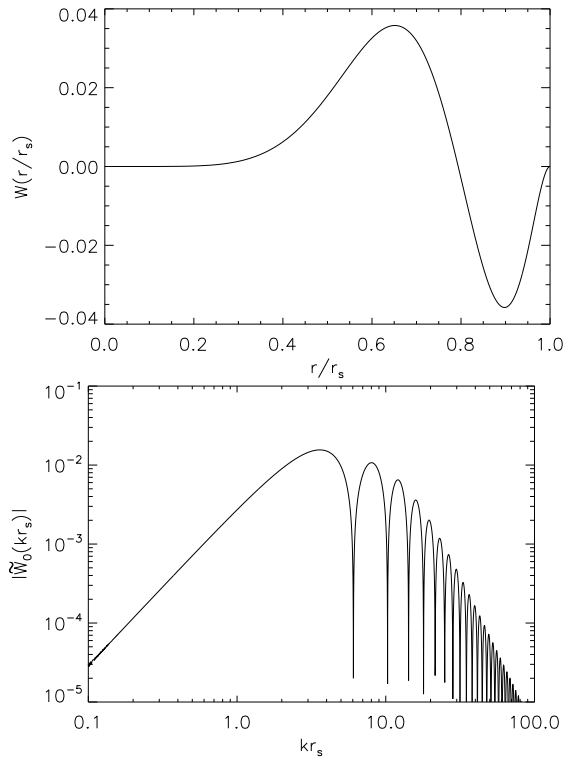

\plotone{rfilter.epsi}
\plotone{kfilter.epsi}
\caption{
(top) The filter $W_\ell(r/r_s)$ in configuration space.
There is a broad hump matching the width of the acoustic peak in
$\xi(r)$ which peaks at $r\simeq 0.65\,r_s$ and a sharp (negative)
spike at $0.9\,r_s$ with a width on the order of $10\%$.  A filter of
this shape will smear a feature, such as the acoustic peak in $\xi(r)$,
by only a small amount which means that the acoustic information will
be well localized in $\wl$. Its compensated nature implies that $\wl$
is not sensitive to the integral constraint.  (bottom) The filter
$\widetilde{W}_\ell(kr_s)$ for $\ell=0$.  The insensitivity of this
filter to large scales is reflected in the fact that it is singly
compensated and $\widetilde{W}_\ell\sim k^2$ as $k\to 0$.  At small scales
$\widetilde{W}_\ell(kr_s)\to \cos(kr_s)/(kr_s)^4$. This is a much more
rapid drop-off than observed in the kernel for $\xi(r)$, which scales as
$(kr)^{-1}$.  These properties of the filter imply that $\wl$ only probes
a narrow range of scales in Fourier space, and that it is insensitive
to large scale fluctuations or poorly constrained small-scale structure.
\label{fig:filt}}
\end{figure}

Following \citet{Pea07} we consider a low order, smooth compensated
filter.  For simplicity we assume $W_\ell$ is independent of
$\ell$, though we could of course choose different weights for each
multipole\footnote{For example, we could make the $k$-weight for $\ell=0$
and $\ell=2$ equal. Doing so facilitates the computation of 
$Q(r) = \frac{\xi_2(r)}{3/r^2 \int_0^r \xi(r')r'^2dr' - \xi_0(r)} = \frac{4/3\beta +  4/7\beta^2}{1+2/3\beta+(1/5)\beta^2}$ 
\citep{Hamil92}, when transformed to $r_s$ space, since it involves both
the monopole and the quadrupole. This ratio is useful
for estimating the anisotropic parameter $\beta$, however, it has limited
applicability outside linear theory.}.
In terms of $x\equiv (r/r_s)^3$, the filter
\begin{equation}
  W(x) = (2x)^2(1-x)^2\left(\frac{1}{2}-x\right)\frac{1}{r_s^3}
\label{eqn:wldef}
\end{equation}
satisfies $W(0)=W'(0)=W(1)=W'(1)=0$ and $\int\ dx\,W(x)=0$.  The suggested
form in configuration space (top panel of Figure \ref{fig:filt})
has a broad hump peaking at $r\simeq 0.65\,r_s$ that matches the
width of the acoustic peak in $\xi(r)$ and a sharp (negative) spike
at $0.9\,r_s$ of width ${\mathcal O}(10\%)$.  This filter will smear
a feature, such as the acoustic peak in $\xi(r)$, by very little
which means that the acoustic information will be localized in $\wl$,
however, not as localized as in $\xi(r)$.  Obviously, given sufficient
signal-to-noise, measuring $\omega_\ell(r_s)$ for many $r_s$ values would
allow resolution in $\xi(r)$ even below the intrinsic width of $W(r/r_s)$
(see \S\ref{sec:om_cov}).

With this choice of $W_\ell$ the window function $\widetilde{W}_\ell$
can be computed analytically (see Appendix for numerical
details) or numerically via fast Hankel transforms.  We show
$\widetilde{W}_\ell(kr_s)$ for $\ell=0$, in the bottom panel of
Figure \ref{fig:filt}.  Since the filter is singly compensated,
$\widetilde{W}_\ell\sim k^2$ as $k\to 0$, reflecting insensitivity
to large scales.  At small scales $\widetilde{W}_\ell(kr_s)\to
\cos(kr_s)/(kr_s)^4$, a much more rapid convergence than evinced by the
kernel for $\xi(r)$, which scales as $(kr)^{-1}$.  Thus $\omega_\ell$
probes a narrow range of scales in Fourier space and is insensitive to
fluctuations on large scales or poorly measured or modeled small-scale
structure.  One can choose the range of $k$ to be sampled by appropriate
choice of $r_s$: more information from high $k$ modes can be included
by using smaller $r_s$.

\begin{figure}
\plotone{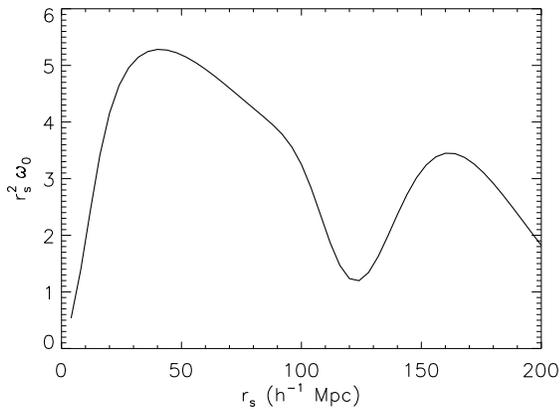}
\caption{Linear theory monopole statistic $\wnot$. The acoustic information
can be seen around the acoustic scale, mainly localized into a single dip 
feature.
\label{fig:linom}}
\end{figure}

As an example, the linear theory monopole statistic $\wnot$ is plotted in
Figure \ref{fig:linom}. Plotting $r_s^2\omega_0$ versus $r_s$ gives a
convenient vertical range. The acoustic information is mostly localized 
into a single dip around the acoustic scale (see \S\ref{sec:locate}).

\subsection{Computation}\label{sec:omegacomp}
It is possible to adapt $\wl$ into a sum over unbinned pair counts for any
sample of mass tracers following the methods described in \citet{Pea07};
there is no need to first compute $\xi(r,\mu)$ via binning of data.
Pair counting allows us to record each galaxy's angle to the LOS ($\mu$)
accurately. Hence, like $\xi(r)$, $\wl$ is sensitive to any anisotropies
of the tracer distribution in clusters (discussed in \S\ref{sec:intro}).

The redshift-space correlation function can be estimated as
\begin{equation}
\label{eqn:xiest}
\xi_s(r,\mu) = \frac{DD(r,\mu)}{RR(r,\mu)} -1 ,
\end{equation}
where $DD(r,\mu)$ is the number of data tracer pairs separated by $r$
and have LOS angle corresonding to $\mu$. $RR(r,\mu)$ is the analogue
for randomly distributed points, normalized to the data counts by a
factor of $N_D^2/N_R^2$. Here, $N_D$ and $N_R$ are the total number of
data and random points respectively.  When analyzing observational data,
the number of random points needs to be much larger than the number of
data points to keep the shot-noise in $RR$ smaller than that in $DD$,
especially at small $r$.  For simulation data, however, it is not
necessary to use a very large number of random points to compute $RR$
smoothly at small scales (elicited below).

Equation (\ref{eqn:xiest}) implies that Equation (\ref{eqn:omxidef}) can
be rewritten as
\begin{equation}
\label{eqn:omxiest}
\omega_\ell(r_s) = i^\ell \int d^3 r W_\ell(r)L_\ell(\mu)
\frac{DD(r,\mu)}{RR(r,\mu)} .
\end{equation}
The $-1$ integrates to 0 due to the compensated nature of the filter.

The $RR$ piece is purely geometrical and is dependent only on the
survey geometry (encoded in $\Phi(r,\mu)$) and the number of random 
points. Hence we can write $RR$ as
\begin{equation}
RR(r,\mu)=2\pi n_DN_D r^2 \Phi(r,\mu) dr d\mu
\end{equation}
where $n_D$ is the number density of data points which is easily
calculable for surveys with well defined boundaries. The above
equation defines $\Phi(r,\mu)$ to be any mismatch between infinite
sized surveys/simulations and finite sized ones due to the presense
of boundaries. The $n_DN_D$ factor is due to the normalization of
the RR counts as mentioned above.  For observations, $\Phi(r,\mu)$
can be computed via binning methods and then fit using a smooth
function $\hat{\Phi}(r,\mu)$. Note that the binning mentioned here
is only required in the computation of the $RR$ counts; there is no
need to bin the data. In the case of simulations in a periodic box,
$\hat{\Phi}(r,\mu)$ is constant as the volume is effectively 
infinite.

With these points in mind, we can now pick arbitrarily small
bins when computing the $DD$ counts since $RR$ has been approximated by a
smooth function and hence does not suffer from shot-noise induced through
pair-counting. As is such, we may employ a binning scheme in which there
is either zero or one $DD$ pair per bin.  This step reduces the integral
in Equation (\ref{eqn:omxiest}) to a sum over $DD$ pairs as in
\begin{equation}
\omega_\ell(r_s)=i^\ell 
\sum_{i \in DD} \frac{W_\ell(r_i)L(\mu_i)}{n_DN_D V \hat{\Phi}(r_i,\mu_i)} .
\end{equation}
Since the estimator can be written as a summation, there is no longer
a need to bin data at all.

\subsection{Covariance Matrix}\label{sec:om_cov}
Since $\omega_\ell$ does not require the binning of data, we can in principle
estimate it at as many $r_s$ values as we wish without affecting the
signal in the adjacent values: there is no bin which is made smaller.
However adjacent points become increasingly correlated as the $r_s$
spacing decreases, compromising the usefulness of very fine sampling.

In the Gaussian limit, the covariance matrix is
\begin{eqnarray}\label{eqn:cov}
  {\rm Cov}&&\left[ \omega_\ell(r_s) , \omega_{\ell'} (r_s') \right] = \frac{2(2\ell+1)(2\ell'+1)}{V} \nonumber \\
  &&\times \int\frac{k^2dk}{2\pi^2}
  \widetilde{W}_{\ell}(kr_s) \widetilde{W}_{\ell'}(kr_s')
  {\mathcal I}_{\ell\ell'}(k)
\end{eqnarray}
with
\begin{equation}
  {\mathcal I}_{\ell\ell'} = \frac{1}{2}\int d\mu
  L_\ell(\mu)L_{\ell'}(\mu) \left[ \sum_L P_L(k)L_L(\mu)+\aleph\right]^2
\end{equation}
where $\sum P_L(k)L_L(\mu)$ is the legendre decomposition of the full 
3D power spectrum $P(\vec{k})$ and $\aleph$
is shot-noise. Assuming Poisson shot-noise, $\aleph = \bar{n}^{-1}$, 
where $\bar{n}$ is the number density of the mass tracer.

\subsection{Summary of Key Features}\label{sec:om_key}
We conclude this section with a summary of the key features and 
advantages of $\wl$ over $\xi(r)$ and $P(k)$ estimators.
\begin{enumerate}
\item
$\wl$ has a compensated filter that reduces sensitivity to poorly
constrained large scale power and hence the integral constraint.
$\xi(r)$, on the other hand, experiences these problems. The compensated
filter also makes $\wl$ measured in different subvolumes of the survey
more independent which is important for attaining robust error estimates
from methods such as bootstrap and jackknife.
\item
The filter is approximately compact in both configuration and Fourier
space. The smoothness in configuration space leads to the steep drop-off
at high $k$ in Fourier space. This effectively minimizes the impact
of large $k$ or small scale power which is not well constrained in
large cosmological surveys.  The filter is localized in configuration
space which means that, unlike in $P(k)$, the acoustic information
is localized in $\wl$. However, it is not as localized as in $\xi(r)$
(see \S\ref{sec:locate}).
\item
Like $\xi(r)$, $\wl$ can be easily adapted into a pair count statistic,
so the relative positions of tracers and each pair's angle from the LOS
is accurately recorded. Hence, it estimates the underlying galaxy
distribution without the need to deconvolve a window function (as
in the case of $P(k)$) and is sensitive to any anisotropies in the
clustering of tracers. The anisotropies can be used to determine the
underlying cosmology.  $P(k)$ estimators do not typically allow this
type of analysis.
\item
There is no need to bin data when computing $\wl$, unlike when estimating
$\xi(r)$ and $P(k)$. 
\end{enumerate}

While the setbacks of traditional $P(k)$ and $\xi(r)$ estimators are 
minor, it is beneficial to have a statistic like $\wl$ which combines many 
of the advantages of both. 

\section{Simulations and Analysis Methods}\label{sec:simmeth}
A major goal in developing this new statistic is to use it for better
measuring and calibrating the acoustic scale. The monopole statistic
$\wnot$ is especially useful in this regard as it is a direct map from the
traditional 2-point correlation function and power spectrum. Therefore,
any results obtained from BAO analysis via these three statistics
is readily comparable. As with the quadrupole of the power spectrum
$P_2(k)$ \citep{PW08}, anisotropic BAO analysis can be performed using
the quadrupole $\omega_2(r_s)$. However, as the main goal of this paper
is to give a broad introduction to the $\wl$ statistic and a simple
demonstration of its application, we defer detailed discussion of
$\omega_2(r_s)$ and anistropic BAO to a future paper.

In this section, we implement $\wnot$ through the use of pure dark matter
N-body simulations. It is possible to model a variety of different galaxy
populations and biases through application of appropriate HODs to the
halos found in the simulations. This is important in demonstrating
the robustness of the $\wl$ statistic over a diverse set of galaxy
populations. We compute $\wnot$ for dark matter and the mock galaxy
populations created via the HODs, and demonstrate how it can be used
to measure the shift in the acoustic peak. This quantity is important
in constraining the precise size of the acoustic scale, which may be
slightly different from that predicted by linear theory due to nonlinear
structure growth. In order to quote the shift with accurate errors, we
use a resampling technique described in \S\ref{sec:fitsamp} which gives us
a large number of shifts from which to calculate a mean and a standard
error of that mean.

\subsection{Simulations}
Our simulations were performed using a code developed by Metchnik and
Pinto which employs a new method \citep{MP10} to compute N-body forces
under periodic boundary conditions. Rather than resorting to the Fourier
methods of PM schemes or using Ewald sums to represent periodicity,
this new method represents periodicity directly, expressing the force on
a particle as due to the rest of the simulation volume and an infinite
sum over its periodic images.

This is made more efficient by partitioning the computational domain into
a three-dimensional grid. The acceleration on particles within a grid
cell is divided into two parts: a near field and a far field. The near
field is that due to the other particles in the cell and to particles in
the adjacent 26 cells. In these calculations, the near-field acceleration
was computed using the direct, $O(N^2)$ method, with Plummer softening.

The far field acceleration on particles in the cell due to each more
distant cell on the grid is represented as a Taylor series expansion
based on the multipole moments in the distant cell. The contribution
from all periodic images of the distant cell is included by recognizing
that the multipole moments in a cell are identical to those in all
of its images. Thus, the sum over images depends only on the (fixed)
geometry of the grid and need be performed only once. The (relatively)
small set of values which results provides a simple and rapidly-evaluated
relation between the multipole moments in distant grid cells and the
Taylor coefficients of the expansion for the acceleration in a given
cell, all under periodic boundary conditions. The acceleration due to
all of the periodic images of the cell and its 26 neighbors is included
in a similar manner.

The calculations described here used order-16 expansions, providing an
overall force accuracy per particle which agrees with Ewald summation
to better than six decimal digits. Overall, the method is significantly
faster and requires significantly less memory than other methods for
computing accelerations in large, periodic N-body problems, while
providing accelerations to machine precision for all particles. Time
integration was performed using a standard kick-drift-kick algorithm,
using independent time steps for each particle.

We derive our results in redshift-space at $z=1$ from a set of 44
simulations with $1024^3$ dark matter particles in each. The simulations
were performed in 1$h^{-1}\Gpc$ periodic boxes with the WMAP5+SN+BAO
best-fit cosmological parameters: $\Om=0.279$, $\OL=0.721$, $h=0.701$,
$\Ob=0.0462$, $n_s=0.96$ and $\sigma_8=0.817$ \citep{Kea09} which implies
a particle mass of $7.2 \times 10^{10} h^{-1} M_\odot$. The initial
conditions are generated via the second-order Lagrangian perturbation
theory code of \citet{Sirko05} at $z=50$ with no extra power for the
box scale.

\subsection{Halo Occupation Distributions}
We use a simple friends-of-friends \citep{Dea85} method with a
linking length equal to $0.16$ of the interparticle spacing to identify the 
collapsed dark matter halos in our simulations. We then populate these 
halos with galaxies by applying simple HODs based on the form
\begin{equation}\label{eqn:halo}
\langle N_g(M) \rangle = [1+(M/M_{\rm sat})^\gamma]\mathrm{exp}(-M_{\rm cen}/M)
\end{equation}
where $M$ is the halo mass, $M_{\rm cen}$ is the minimum mass for a halo to
contain a central galaxy, $M_{\rm sat}$ is the minimum mass for a halo to 
contain at least one satellite, and $\gamma$ is an exponential parameter that we
set to $1$ \citep{GS02, Bea03, Kea04, Zea05}. We assign a central galaxy to 
a halo if $M > M_{\rm cen}$, this is a good approximation to
$\langle N_{\rm cen}(M) \rangle = \mathrm{exp}(-M_{\rm cen}/M)$. The
central galaxy is taken to be at the halo's center of mass and assigned the
center of mass velocity.
If a halo is assigned a central galaxy, then the number of satellite galaxies
is determined by generation of a random
integer based on a Poisson distribution with mean equal to
$\langle N_{\rm sat}(M) \rangle = (M/M_{\rm sat})^\gamma$. We then randomly 
pick a corresponding number of halo particles and assign their positions and
velocities to the satellites.

In order to compare the peak shifts derived using $\wnot$ and $P(k)$ in a 
range of models, we apply three different HODs to our simulations.
The properties of these HODs are described in Table \ref{tab:hod_p} and
obtained by adjusting the values of $M_{\rm cen}$ and $M_{\rm sat}$. We also 
list the properties of a dark matter (DM) only case that is merely a $0.4\%$
subsample of the particles in each simulation. We compute $\wnot$ in 
redshift-space via the pair counting method detailed in \S\ref{sec:omegacomp} 
within the range $5h^{-1}\Mpc \leq r_s \leq 200h^{-1}\Mpc$ using $5h^{-1}\Mpc$ 
spacing. We also compute $\sigma_8$, the RMS mass fluctuation within a 
$8h^{-1}\Mpc$ radius, using a similar pair counting method derived from the 
configuration space equation for a general radius R 
(\citet{Zehavi05})
\begin{equation}
\sigma_R^2 = \int^{2R}_{0}\frac{1}{R^3} \left[ 3 - \frac{9}{4} 
\frac{r}{R} + \frac{3}{16} \left( \frac{r}{R}\right)^3 \right] r^2\xi(r)dr.
\end{equation}

\subsection{Fitting $\wnot$ to Measure the Peak Shift}\label{sec:fittech}
We fit the redshift-space $\wnot$ using the form
\begin{equation}\label{eqn:fitform}
P_{\rm sim}(k) = B(k)P_{m}(k/\alpha) + A(k)
\end{equation}
where
\begin{equation}
B(k) = \frac{(b_1^2+b_2^2k+b_3^2k^2)}{1+r_{\rm scale}k}
\label{eqn:fitb}
\end{equation}
and $A(k)$ transforms into $A(r_s)=a_1r_s^{-9}$ in $r_s$
space. Such a form for $A(k)$ is motivated by the fact that we
want to marginalize over the shape of the correlation function at
small scales (i.e., the contribution of the 1-halo term). Expanding
Equation (\ref{eqn:wldef}), we see that 
$W(x) = (-4x^5 + 2x^4 - 8x^3 + 2x^2)\frac{1}{r_s^3}$. 
This implies that the highest order term in $r_s$ is $r_s^{-9}$. At
small scales, this is the term that will dominate in the 
transformation from $\xi(r)$ to $\wnot$ as defined by Equation
({\ref{eqn:omxidef}}). 

$P_m(k)$ is the template power spectrum we use for our fitting. To
account for the degradation of the acoustic peak through nonlinear
evolution and redshift distortions, the template model $P_m(k)$ is
obtained from the linear power spectrum $P_{\rm lin}(k)$ at $z=1$ by
the modification
\begin{equation}
P_m(k) =
[P_{\rm lin}(k)-P_{\rm smooth}(k)]\mathrm{exp}(-k^2\Sigma_{\rm nl}^2/2)+
P_{\rm smooth}(k) \label{eqn:apodize}
\end{equation}
where $P_{\rm smooth}$ is the dewiggled power spectrum described in
\citet{EH98} and $\Sigma_{\rm nl}$ is a nonlinear parameter used to
degrade the peak \citep{Eea05, Tea06, CS06, Eea07, CS08, M08}. To
allow maximum flexibility in our marginalization, we marginalize over
$\Sigma_{\rm nl}$ and the other nonlinear nuisance parameter in
equation (\ref{eqn:fitb}), $r_{\rm scale}$.

Our scale dilation parameter $\alpha$ represents the shift in
the acoustic peak. Under this formalism, $\alpha > 1$ indicates a
shift towards smaller scales and $\alpha < 1$ indicates a shift
towards larger scales. Physically, $\alpha$ is the ratio between
the linear theory acoustic scale (150 Mpc) to the measured
acoustic scale. Since all the terms in the fitting function are
additive, the basis functions $P_m(k/\alpha)/(1+r_{\rm scale}k)$,
$kP_m(k/\alpha)/(1+r_{\rm scale}k)$, $k^2P_m(k/\alpha)/(1+r_{\rm
scale}k)$, and $r_s^{-9}$ for a fixed $\alpha$, $\Sigma_{\rm nl}$ and
$r_{\rm scale}$ can be easily mapped into $r_s$ space (if necessary)
using equation (\ref{eqn:omdef}). A least-squares fit using the mapped
basis functions is then performed against $\wnot$ from the simulations
to obtain values for the linear nuisance parameters $b_1$, $b_2$, $b_3$
and $a_1$.  As we are interested in the acoustic feature, we use a fitting
range of $30 \leq r_s \leq 200 h^{-1}\Mpc$. For an $r_s$ spacing of $5
h^{-1}\Mpc$, this implies 28 degrees of freedom in the fit, where the
number of degrees of freedom is defined as the difference between the
number of data points being fit and the number of parameters in the
fitting form.

We assume that the errors on $\wnot$ can be well approximated
by the covariance matrix $C$ assuming Poisson shot-noise (see
\S\ref{sec:fitsamp}) with the addition of nonlinear shot-noise (see
equation (\ref{eqn:nlshot}) and surrounding text). We also assume that
the monopole ($\ell=0$) dominates $P(\vec{k})$ so that all higher order
contributions to the power spectrum are effectively zero. This amounts to
computing $C$ using $P(\vec{k})=P_m(k)$ as the input power spectrum, where
we take a fixed $\Sigma_{\rm nl}=7.0h^{-1}\Mpc$ at $z=1$ in redshift-space
following \citet{Sea08}. We normalize this power spectrum to the amplitude
of the redshift-space power spectrum through multiplication by the bias
squared defined initially as $b^2 = (\sigma_{\rm 8,case}/\sigma_{\rm
8,matter})^2$. The values for $\sigma_{\rm 8,case}$ are given in Table
\ref{tab:fit} and $\sigma_{\rm 8,matter}=0.506$ in real space at $z=1$
in linear theory.  We want to ensure that the input power spectrum
to the covariance matrix calculation is as close as possible to the
simulation data so that the covariance matrix is a reliable estimate of
the errors. To do this, we marginalize over the average $\wnot$ of the
44 simulations for each case to obtain a value for the leading order
term that scales $P_m(k)$: $b_1^2$ in the fitting form of equation
(\ref{eqn:fitb}). We then iterate this marginalization and scale $b^2$
by the values of $b_1^2$ obtained until the output $b_1^2$ from this
iterative fitting is close to $1$.  We expect that scaling $P_m(k)$
by this final value of $b^2$ will approximate the simulation data well
and hence be valid input to the covariance matrix calculation for the
resampling techniques described in \S\ref{sec:fitsamp}.

The shot-noise we enter into the calculation of $C$ includes a nonlinear
component (quoted in Table \ref{tab:hod_p}) in addition to the Poisson 
shot-noise $\bar{n}^{-1}$ as described in Equation (\ref{eqn:cov}). This 
additional shot-noise is a result of nonlinear structure formation on small 
scales. We estimate this nonlinear shot-noise as
\begin{equation}
\aleph_{\rm nonlin} = \int_0^{r_{\rm nonlin}}4\pi r^2[\xi(r)-\xi_{\rm lin}(r)]dr
\label{eqn:nlshot}
\end{equation}
where $\xi(r)$ is the correlation function averaged over the $44$
simulations for each HOD, $\xi_{\rm lin}(r)$ is the linear correlation
function at $z=1$ and $r_{\rm nonlin}$ is the scale above which nonlinear
effects become unimportant. We take $r_{\rm nonlin}$ to be $10h^{-1}
\Mpc$. The resulting $\aleph_{\rm nonlin}$ is a rough estimate of the excess
small scale correlation due to nonlinear evolution. Since it makes little
difference whether all of this extra shot-noise comes in at zero 
separation in $r$ or through the extended effects of the one-halo term,
which is only important at small $r$, we assume the excess correlation
to be a spike at $r=0$ for convenience. When transformed into $k$
space, this gives a constant and becomes extra white noise that we
add on to every mode equally, in addition to the Poisson shot-noise.
For the subsampled DM case, the linear and the measured correlation
functions were sufficiently similar at $1$-$10h^{-1}\Mpc$ to warrant
taking $\aleph_{\rm nonlin}=0$ for this case.  Alternatively, one can
also account for nonlinear shot-noise by computing the covariance matrix
using the nonlinear power spectrum, but this is more computationally
challenging.

The $\chi^2$ likelihood indicator corresponding to the best-fit
linear nuisance parameters for fixed $\alpha$, $\Sigma_{\rm nl}$ and
$r_{\rm scale}$ is then
\begin{equation}\label{eqn:chi2}
\chi^2 = (\vec{\omega}_0-\vec{m})^TC^{-1}(\vec{\omega}_0-\vec{m})
\end{equation}
where $\vec{\omega}_0$ is $\wnot$ measured from the simulations,
$\vec{m}$ is the best-fit model and $C^{-1}$ is the inverse of the
covariance matrix. We compute the best-fit values of $\alpha$,
$\Sigma_{\rm nl}$ and $r_{\rm scale}$ by minimizing $\chi^2$ of the fits
for the DM case and for each HOD using a generalized reduced gradient 
method from IDL. We quote the bias for each case as 
$\sigma_{\rm 8,case}/\sigma_{\rm 8,matter}$ 
multiplied by the additional scaling factors of $b_1$ described above.

\begin{figure}
\plotone{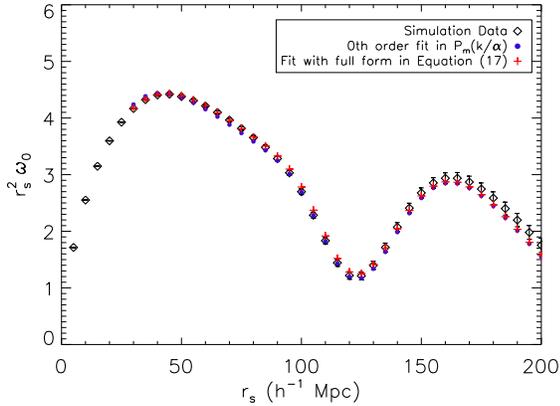} \caption{$\wnot$ averaged over all $44$
simulations for HOD$1$ (black diamonds). Overplotted are a fit
obtained through the form in Equation (\ref{eqn:fitform}) (red
crosses) and a $0th$ order fit in $P_m(k/\alpha)$ (purple dots),
both over a range of $30\le r_s\le 200 h^{-1}\Mpc$. One can see that
the $0th$ order fit already appears quite good with $\chi^2=2.20$ per dof.
However, by introducing additional nuisance parameters, the quality of 
the fit over the specified range improves further to $\chi^2=0.94$ per dof.} 
\label{fig:fit}
\end{figure}

We plot $\wnot$ averaged over all $44$ simulations for HOD$1$ in Figure
\ref{fig:fit}. Overplotted are the marginalization obtained through the
form in Equation (\ref{eqn:fitform}) and a $0th$ order fit obtained by
just a rescaling of $P_m(k/\alpha)$, i.e. $b^2P_m(k/\alpha)$ where $b$
is the only fit parameter.  Although the $0th$ order fit already appears
quite good, one can see that introducing additional nuisance parameters
improves the quality of the fit even more over the fitting range. The
$\chi^2$ per degree of freedom (dof) improves from 2.20 to 0.94.

Using this fitting technique, we derive values of bias, $\alpha$
and associated errors for each of our three HODs and our DM case
via the resampling methods described in the following section.

\subsection{Resampling Methods}\label{sec:fitsamp}
We use two different methods to measure the mean peak shift $\alpha$
and the scatter in the mean $\sigma_\alpha$ for each case in Table
\ref{tab:hod_p}.  The first is a modified jackknife technique in which
we randomly select M out of N simulations at a time without replacement,
average their $\wnot$ and fit this average.  We repeat this 1000 times
and extract an average $\alpha$ and a scatter in $\alpha$. This scatter
needs to be rescaled by an additional factor of $f=\sqrt{M}/\sqrt{N-M}$
in order to reflect the scatter asscociated with the mean of $\alpha$ for
N simulations. For our simulations we have $N=44$ and take $M=22$. With
this choice of $M$, $f=1$ and so the scatter in $\alpha$ reflects the
error in the mean of $\alpha$. This method is useful in that it provides
us with a large set of $\alpha$'s from which we can accurately derive
a mean $\alpha$ and $\sigma_\alpha$.

The fit results for the subsampled DM case as well as for each HOD model
are quoted in Table \ref{tab:fit}. The average values of $\Sigma_{\rm nl}$
and $r_{scale}$ are also included for completeness, however the focus
of this paper is on $\alpha$.  The values of $\alpha$ we obtain are
$1\sigma$ consistent with those derived using the perturbation theory
results of \citet{PdW09}.  The somewhat low value of $\chi^2$ per dof
for HOD$3$ suggests that we are overestimating the amount of nonlinear
shot-noise. We are also approaching the shot-noise limited regime for
HOD$3$ as evidenced by the fact $\bar{n}P_{0.2}\approx1.6$. We note here
that we used $\chi^2$ only to find the best-fit $\alpha$ for each HOD, not
to generate the errors. Hence, the fact that our reduced $\chi^2$ values
are slightly deviant from unity does not hinder the error estimation.

The second method we use is jackknife resampling. The results obtained
using this method are in good agreement with the first method. This 
indicates that the error estimates obtained using our first method are
robust in comparison to more traditional methods.

By using these resampling techniques, any non-Gaussian effects not
accounted for by assuming a Gaussian covariance matrix while fitting
(as in \S\ref{sec:fittech}) will be reflected in $\sigma_\alpha$.

\section{Comparison to the Power Spectrum}\label{sec:comppk}

\subsection{Comparison from Simulations}
\begin{figure}
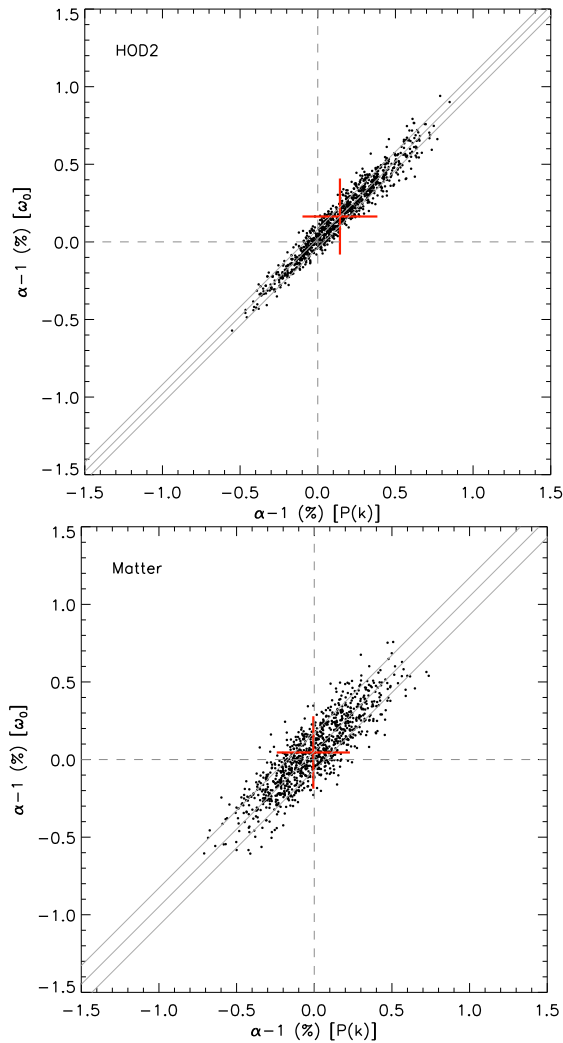

\plotone{n1e6.sat5.epsi}
\plotone{matter.epsi}\\
\caption{$\alpha$ from $\wnot$ versus $\alpha$ from $P(k)$ for
HOD$2$ (top) and DM (bottom). The data points
are from a resampling technique in which we randomly pick $M=22$
simulations out of $N=44$ total and fit the averaged $\wnot$ from
these $M$ simulations. We repeat this $1000$ times and hence
obtain $1000$ values of $\alpha$. The scatter on $\alpha$ needs to
be rescaled by $\sqrt{M}/\sqrt{N-M}$ to reflect the true scatter
on the mean. For our choice of $M$, this scaling factor is equal
to $1$. Hence the scatter in the plot truly reflects the scatter
on the mean of $\alpha$. The red cross marks the mean $\alpha$
values with their associated errors. The central grey line has unity
slope and passes through the mean. The two outer
grey lines delineate the $1\sigma$ boundaries associated with
$\Delta\alpha_{\omega P}$. As the data points lie largely in between 
the $1\sigma$ lines with a slope similar to unity for both HOD$2$ and 
the DM case (see Table \ref{tab:aldiff}), we conclude that the two 
$\alpha$ sets are consistent with each other. The same correlation is 
observed for HOD$1$. HOD$3$ shows
$1.6\sigma$ agreement between $\wnot$ and $P(k)$. This slightly larger 
discrepancy may be due to the fact that shot-noise is becoming significant 
in this low number density case. Also, shot-noise may affect $P(k)$ and 
$\wnot$ differently or the nuisance parameters may not be fully 
handling the scale-dependence of a high-bias HOD such as HOD$3$. The
large scatter in the DM case is likely due to the subsampling of matter
in the computation of $\wnot$ but not in $P(k)$. The overall agreement 
between the $\wnot$ and $P(k)$ results imply that distance measures
will be consistent between the two.
\label{fig:omvpk}}
\end{figure}

An important step in implementing this new statistic is to show that it
produces consistent results when compared to established methods and
can therefore be an effective calibrator of the acoustic scale.
We do this by comparing the peak shifts measured from the same set of
simulations via the new $\wnot$ statistic and the traditional $P(k)$
method (see S\&M for details).

As we use the same 1000 random sets of M simulations as S\&M, there should
be a 1:1 correspondance between the $\alpha$'s derived from $\wnot$
and $P(k)$ for the DM case and for each HOD. It should be noted that
S\&M use different $P(k)$ fitting forms from the one detailed in
\S\ref{sec:fittech}. They employ two fitting forms, both of which can
also be described by equation (\ref{eqn:fitform}). The first form has
$B(k)$ as a $2^{\rm nd}$ order polynomial and $A(k)$ as a $7^{\rm th}$
order polynomial. The second form uses Pade approximants for $B(k)$,
i.e., $B(k)=b_0(1+c_1k+c_3k^2+c_5k^3)/(1+c_2k+c_4k^2)$ and a $2^{\rm
nd}$ order polynomial for $A(k)$. We have chosen a different form in
this work to induce better convergence of the integral from Fourier
space to $r_s$ space while transforming the basis functions. We compare
the $\alpha$'s measured from $\wnot$ against those measured from $P(k)$
by \citet{Sea10} (DM) and \citet{Mea09} (HODs) using the first form. It
should also be noted here that the $P(k)$ results obtained for the DM case
by \citet{Sea10} utilize the full DM sample whereas we have subsampled
to reduce computation time in this work. The methodology used to derive
the $P(k)$ results are described in detail in \citet{Sea10}.

Figure \ref{fig:omvpk} shows $\alpha$ from $\wnot$ versus $\alpha$ from
$P(k)$ for the 1000 fit iterations performed on HOD$2$ (top) and DM
(bottom). The red cross indicates the mean $\alpha$ values with their
associated errors. The central grey line is a line with slope unity that
passes through the mean. The two outer grey lines indicate the $1\sigma$
boundaries associated with $\Delta\alpha_{\omega P} = \alpha_{\omega} -
\alpha_{P}$. The mean difference between $\alpha_{\omega}$ and $\alpha_P$
along with the standard deviation is quoted in Table \ref{tab:aldiff}
under $\langle \Delta\alpha_{\omega P} \rangle$ for the DM case as
well as each HOD case. The plots indicate that the correlation between
$\alpha_{\omega}$ and $\alpha_P$ is $1\sigma$ consistent with a line
of slope unity that has a y-intercept of $0$ for both HOD$2$ and the
DM case. This implies that the two $\alpha$ sets are consistent with
each other. The same holds for HOD$1$ as evidenced through the table,
but HOD$3$ is slightly more deviant with $1.6\sigma$ agreement between
$\wnot$ and $P(k)$. The larger discrepancy between the HOD$3$ results
may be due to the fact that shot-noise is becoming significant in this
case (as shown in \S\ref{sec:fitsamp}). It could also be that shot-noise
affects $P(k)$ and $\wnot$ differently or the nuisance parameters are
not fully handling the scale-dependence of a high-bias HOD such as
HOD$3$. The large scatter in the DM case is likely due to the fact that
we have subsampled the matter in our computation of $\wnot$ but not in
$P(k)$. The $\alpha$'s from most of the cases are $1\sigma$ consistent
between $\wnot$ and $P(k)$, indicating that distance measures will be
consistent between the two statistics. This also indicates that any
systematics introduced by using the different fitting forms for $P(k)$
and $\wnot$ are minor. Hence we conclude that $\wnot$ is a well-tuned
statistic for analysis of BAOs.

\subsection{Theory Constraints on $\sigma_\alpha$}\label{sec:the_op}
\begin{figure}
\plotone{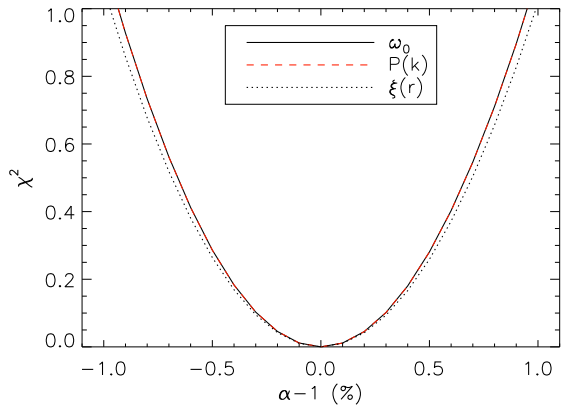} 
\caption{$\chi^2$ versus $\alpha$ for fits in $\wnot$, $P(k)$
and $\xi(r)$. We shift the acoustic feature from linear theory
by a given $\alpha$ (here $\alpha_{given}=1.0$) and run our
fit algorithms to see how well we can recover this input. For
$\wnot$ we fit between $r_s=30$-$210h^{-1}\Mpc$, for $P(k)$ we fit
between $k=0.0$-$1.2h\Mpc^{-1}$ and for $\xi(r)$ we fit between
$r=20$-$200h^{-1}\Mpc$. The parabolic shape of the curves is due to
the fact that $\wnot$, $P(k)$ and $\xi(r)$ are derived from a Gaussian
random field in linear theory. The width of the parabola at $\chi^2=1$
is then the theoretical $\sigma_\alpha$ of the fit.  The $\wnot$ and
$P(k)$ curves overlap nicely, implying that the $\sigma_\alpha$ ratio
between $\wnot$ and $P(k)$ is $\sim1$.  This indicates that for
the given finite fitting ranges, $\wnot$ and $P(k)$ contain equal
amounts of acoustic information (trivially true for infinite fitting
ranges by definition of $\wnot$, however, real surveys are finite in 
size). Hence, the same amount of acoustic information can be obtained
through either $\wnot$ or $P(k)$ analysis of equal volume surveys.}
\label{fig:sigcomp}
\end{figure}

As we wish to promote $\wnot$ as an alternative method for analyzing
the BAO, it is necessary to show how much acoustic information can be
extracted from $\wnot$ relative to $P(k)$ and $\xi(r)$ for surveys of the
same size. If our fitting ranges were infinite, then by the definitions
in \S\ref{sec:omega}, all three estimators should yield the same amount of
BAO information. However, in reality, fitting ranges are finite. 

We investigate the effects of this by shifting the acoustic feature
in the linear theory $\wnot$, $P(k)$ and $\xi(r)$ by a given $\alpha$
and then running our fit algorithms to see how well we can recover this
input $\alpha$. For $\wnot$ we fit between $r_s=30$-$210h^{-1}\Mpc$
in $r_s$ spacings of $2.5 h^{-1}\Mpc$; for $P(k)$ we fit between
$k=0.0$-$1.2h\Mpc^{-1}$ in log($k$) spacings of $\sim0.002$; and
for $\xi(r)$ we fit between $r=20$-$200h^{-1}\Mpc$ in $r$ spacings
of $1 h^{-1} \Mpc$. We use the same fitting technique as described in
\S\ref{sec:fittech} but with different forms for $B(k)$ and $A(k)$ that
are then transformed to $r$ and $r_s$ space to fit $\xi(r)$ and $\wnot$
respectively. This means that the fitting forms for $P(k)$, $\xi(r)$ and
$\wnot$ all derive from the same $B(k)$ and $A(k)$ functions. By enforcing
this consistency between fitting forms, any potential systematics that
may arise due to the use of different fitting forms for each statistic
can be avoided.

We pick $B(k)=b$, where $b$ is the large-scale bias, and $A(k)$ to be
the cold dark matter-only power spectrum multiplied by a set of cubic
spline functions specified at $k=0.01$-$1.09h\Mpc^{-1}$.  The spline
points are picked so that seven of them are logarithmically spaced in
the range $k=0.01$-$0.25$ and seven of them are linearly spaced in the
range $k=0.25$-$1.09$ giving a total of 14 spline points.  The basis
functions specified by the small $k$ spline points are necessary to
allow flexibility in the marginalization of $\wnot$ at large scales.
The derivative is also specified at the first spline point to derive an
additional spline function. The spline functions are taken to be natural
(i.e., second derivative equal to $0$) at the first and last spline
points, beyond which linear extrapolations are implemented. This choice
of $A(k)$ ensures convergence when transformed to $r_s$ space and makes
the fits in $P(k)$, $\xi(r)$ and $\wnot$ readily comparable.  We assume
a survey volume of $1h^{-1}\Gpc$ with $\aleph = 1000$ (i.e., one million
particles). As a cross check, we confirmed that this new fitting form
does in fact give similar results to the form used in \S\ref{sec:fittech}.

Figure \ref{fig:sigcomp} plots the $\chi^2$ versus $\alpha$ for $\wnot$,
$P(k)$ and $\xi(r)$. Here the input $\alpha$ is equal to $1$. If $\wnot$,
$P(k)$ and $\xi(r)$ are derived from a Gaussian random field as they
are in linear theory, we would expect that $\chi^2$ versus $\alpha$
be parabolic as shown in the figure. The width of the parabola at
$\chi^2=1$ is then the theoretical $\sigma_\alpha$ from the fit. The
overlap between the $\wnot$ and $P(k)$ curves indicates that the ratio
of $\sigma_\alpha$ for $\wnot$ to $P(k)$ is $\sim1$. This means that for
the given fitting ranges $\wnot$ and $P(k)$ contain equal amounts of
acoustic information. Since volume is proportional to $\chi^2$ which
is proprotional to $\sigma^{-2}$, an important implication is that we
are able to obtain the same amount of acoustic information using either
$\wnot$ or $P(k)$ analysis of equal volume surveys. We emphasize
here that the results presented in this section assume idealized linear
theory forms for $P(k)$, $\xi(r)$ and $\wnot$. In practice, numerous
physical and observational effects distort the measured statistics
from these ideals. However, we expect that the features described in
\S\ref{sec:omega} will reduce the impact of troublesome observational
effects in any BAO analysis using $\wnot$.

\subsubsection{Locating the Acoustic Information}\label{sec:locate}
\begin{figure}
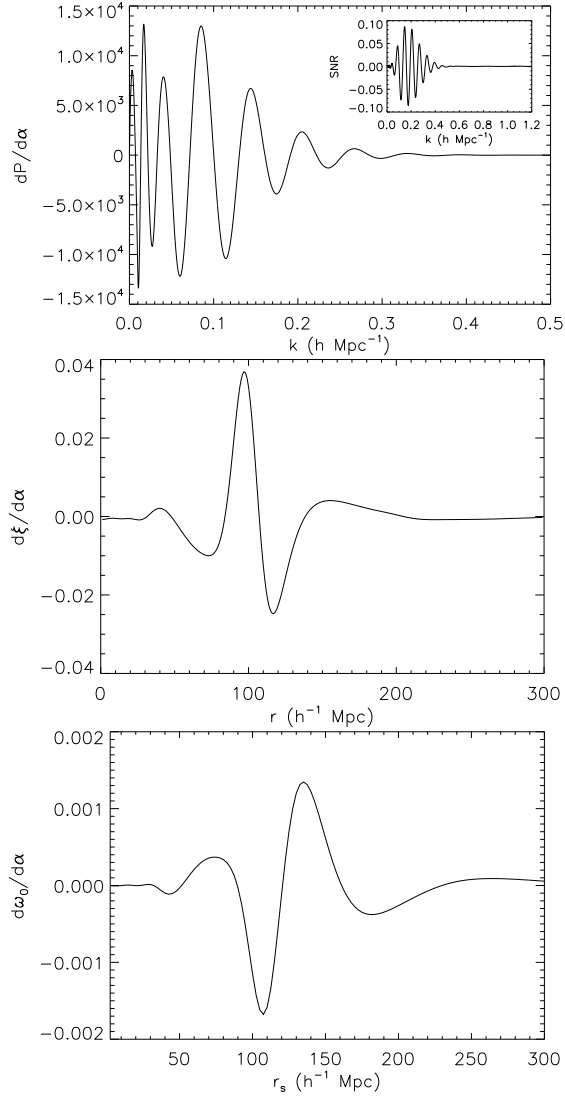

\plotone{dpda_log.epsi} 
\plotone{dxida_log.epsi}
\plotone{domda_log.epsi}
\caption{(top) $dP/d\alpha$ calculated from the residuals of the $P(k)$ 
fits after marginalizing out the broadband shape. The change in $P(k)$
with $\alpha$ captured by $dP/d\alpha$ should correspond to how the
acoustic information is shifted as $\alpha$ changes. The ratio of
$dP/d\alpha$ to $P(k)/k$ approximates the shape of the signal-to-noise
ratio and is shown in the top right-hand corner of the plot. The fact
that the ratio is very small at $k< 0.05 h \Mpc^{-1}$ indicates that
all the ringing in $dP/d\alpha$ at these scales is just noise from
the spline basis functions attempting to match the broadband shape
of $P(k)$ at these scales. Hence, this small $k$ ringing does not
actually reflect the shifting of acoustic information as $\alpha$
changes. The ratio is larger for $k > 0.05 h \Mpc^{-1}$ indicating that
the oscillations in $dP/d\alpha$ at these $k$ truly reflect the shifting
of acoustic information with $\alpha$.  (middle) $d\xi/d\alpha$ obtained
by transforming $dP/d\alpha$. This shows where the acoustic information is
located in configuration space and how it changes with $\alpha$. (bottom)
$d\omega_0/d\alpha$ obtained by transforming $dP/d\alpha$ to $r_s$
space. This shows where the acoustic information is located in $r_s$
space and how it changes with $\alpha$. Comparison with the middle panel
indicates that the acoustic information is not as localized in $\wnot$ as
it is in $\xi(r)$, however, it is still reasonably well localized.  All of
the acoustic information is located within $r_s \sim 30$-$300h^{-1}\Mpc$,
indicating that the optimal fitting range for $\wnot$ is somewhere within
these limits.}
\label{fig:dda}
\end{figure}

It is useful to track down where the acoustic information lies and how it
changes with $\alpha$ in $\wnot$, $\xi(r)$ and $P(k)$. This is reflected
in the derivatives $d\omega_0/d\alpha$, $d\xi/d\alpha$ and $dP/d\alpha$
after marginalizing out the broadband shape. To do this, we calculate
the residuals from the $P(k)$ fits described in \S\ref{sec:the_op}
for $\alpha = 0.996$ and $\alpha=1.004$. These residuals should be
representative of the acoustic signature after the broadband shape has
been marginalized out. We then take $dP/d\alpha$ as the difference between
these residuals divided by $1.004-0.996=0.008$. The transformations of
$dP/d\alpha$ into $r_s$ and $r$ space then give us $d\omega_0/d\alpha$
and $d\xi/d\alpha$ respectively. We have plotted $dP/d\alpha$ in the
top panel of Figure \ref{fig:dda}, $d\xi/d\alpha$ in the middle panel
and $d\omega_0/d\alpha$ in the bottom panel.  If one plots the ratio of
$dP/d\alpha$ to $P(k)/k$, one is left with the approximate shape of the
signal-to-noise ratio \footnote{The noise term $\sigma_P = P(k)/\sqrt{dN}$
where $dN = k^2dk$ is the number of modes out to $k$. For constant
increments in $k$, $dN \propto k^2$ and hence $\sigma_P \propto P(k)/k$}
(SNR). This is plotted in the top right-hand corner of the top panel
in Figure \ref{fig:dda}. One can see that the SNR is small at $k<0.05
h \Mpc^{-1}$, indicating that the small $k$ ringing in $dP/d\alpha$ is
merely noise from the spline basis functions attempting to match the
shape of $P(k)$ at these scales, and is not indicative of the shift
in acoustic information with $\alpha$. The shifting of the acoustic
information with $\alpha$ is only truly evident at $k>0.05 h \Mpc^{-1}$
where the SNR is larger. It is evident from these plots that the acoustic
information is not as localized in $\wnot$ as in $\xi(r)$, but it is still
reasonably well localized.  The bottom panel of Figure \ref{fig:dda}
indicates that the optimal fitting range that will include all of the
acoustic information encoded in $\wnot$ is somewhere within the range
$r_s = 30$-$300h^{-1}\Mpc$.

\begin{figure}
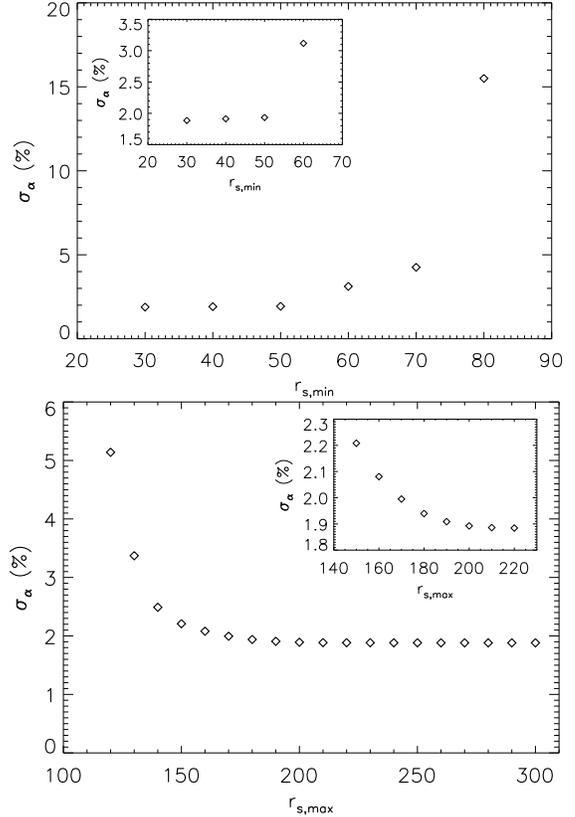

\plotone{rsminvda_new.epsi}
\plotone{rsmaxvda_new.epsi}
\caption{(top) Plots how $\sigma_{\alpha}$ changes as the minimum 
$r_s$ of the fitting range is stepped up from $30$-$80h^{-1}\Mpc$ with
the maximum $r_s$ of the fitting range fixed at $300h^{-1}\Mpc$.
The deviation of $\sigma_\alpha$ at $r_{s,min}$ larger than
$40h^{-1}\Mpc$ indicates that some of the acoustic information is being
missed by these fitting ranges.  (bottom) Plots how $\sigma_{\alpha}$
changes as the maximum $r_s$ of the fitting range is stepped down from
$300$-$120h^{-1}\Mpc$ with the minimum $r_s$ of the fitting range fixed at
$30h^{-1}\Mpc$. The deviation of $\sigma_\alpha$ at $r_{s,max}$ smaller
than $200h^{-1}\Mpc$ indicates that some of the acoustic information is
being missed by these fitting ranges.  The above analysis implies that
in order to extract all of the acoustic information, one needs to fit
between $r_s=40$-$200h^{-1}\Mpc$ at minimum.}
\label{fig:rsvsa}
\end{figure}

The top panel of Figure \ref{fig:rsvsa} shows how $\sigma_{\alpha}$ 
changes as the minimum $r_s$ of the fitting range is stepped up from 
$30$-$80h^{-1}\Mpc$ with the maximum $r_s$ of the fitting range fixed at 
$300h^{-1}\Mpc$. The bottom panel of Figure \ref{fig:rsvsa} shows how 
$\sigma_{\alpha}$ changes as the maximum $r_s$ of the fitting range is
stepped down from $300$-$120h^{-1}\Mpc$ with the minimum $r_s$ of the
fitting range fixed at $30h^{-1}\Mpc$. The regions of the plots where
$\sigma_{\alpha}$ begins to modulate are blown up for clarity. One can see
that when the minimum of the fitting range is larger than $40h^{-1}\Mpc$,
$\sigma_\alpha$ begins to deviate, indicative of missing some of the
acoustic information. This also happens when the maximum of the fitting
range is smaller than $200h^{-1}\Mpc$. Hence, the minimum fitting range
that allows one to extract all of the acoustic information appears to
be $r_s=40$-$200h^{-1}\Mpc$. Hence, to ensure we are encapsulating
all of the acoustic information, we picked the fitting range to be
$r_s=30$-$210h^{-1}\Mpc$ in Figure \ref{fig:sigcomp}.

\section{Conclusions}\label{sec:theend}
We have presented a new statistic $\wl$ for analyzing baryon
acoustic oscillations. This new statistic is advantageous over the
traditional methods used to estimate $\xi(r)$ and $P(k)$ as it
does away with many of their setbacks. Estimators of $\xi(r)$ are
sensitive to poorly measured large scale power through effects such
as the integral constraint, whereas the compensated
nature of the filter $W_\ell(r)$ used to compute $\wl$ circumvents
this problem. We expect that this feature of the filter will also make
$\wl$ measured in different subvolumes of a survey more independent 
which makes error estimation methods such as bootstrap and jackknife more
robust. Estimators of $P(k)$ give the true density field convolved with
a window function making the measured statistic biased.  Attempting to
deconvolve the window function introduces artificial smoothing at small
separations in $k$ due to their near delta function shapes in real
observational surveys. $\wl$, on the other hand, does not suffer from
this as it is a pair count statistic. Pair counting estimators allow us
to record the relative locations of tracer pairs as well as their angles
from the LOS direction. This provides us with an accurate map of the
underlying galaxy distribution. The fact that LOS angle is recorded also
means that any anisotropic clustering should be apparent in $\wl$. Hence
we expect that it can also be used to probe the underlying cosmology. In
addition, there is less need to worry about binning related issues when
computing $\wl$ as we never need to bin the data. The smoothness of the
filter in configuration space causes the rapid fall-off of the filter
in Fourier space. This reduces the impact of large $k$ modes or small
scales which are not well constrained in large cosmology surveys. The
localized nature of $W_\ell(r)$ is conducive to minimal smearing of the
acoustic information so that it is mostly concentrated in a single dip
around the acoustic scale.  This translates to a cleaner representation
of the acoustic information when plotted, as opposed to the oscillatory
features of $P(k)$.

We also showed that with the present form for $W_\ell(r)$ and a
finite fitting range encompassing the acoustic scale, it is possible
to extract the same amount of acoustic information using either
$\wnot$ or $P(k)$ from equal volume surveys.  It is important to
note that these results were obtained through analysis of idealized
linear theory forms for $P(k)$, $\xi(r)$ and $\wnot$. In practice, the
measured forms of these statistics are distorted by various physical and
observational effects. However, we expect that the features described in
\S\ref{sec:omega} will reduce the impact of troublesome observational
effects in any BAO analysis using $\wnot$. We also demonstrated where
the acoustic information is located in $\wnot$, $\xi(r)$ and $P(k)$ and
how it changes with $\alpha$. From this analysis, the minimum fitting
range required to extract all of the acoustic information from $\wnot$
appears to be $r_s=40$-$200h^{-1}\Mpc$.  

We compared the acoustic peak shifts derived using $\wnot$ to those
derived using $P(k)$ for a pure DM case as well as for three halo based
galaxy models.  The results for the DM and the higher number density cases
are all much better than $1\sigma$ consistent with each other.  The low
number density case is slightly deviant with $1.6\sigma$ agreement between
$\wnot$ and $P(k)$.  This may be a result of approaching the shot-noise
limited regime or our lack of understanding of shot-noise in general and
how it may affect $P(k)$ and $\wnot$ differently. It may also be caused by
our fitting form not handling the scale-dependence of high-bias models in
full. The general consistency betwen $\wnot$ and $P(k)$ is encouraging
and implies that distance measures will be consistent between the two
methods. From this and the features listed above, we conclude that $\wnot$
is a well-tuned new statistic for BAO analysis.

\acknowledgements
This work is supported by NASA BEFS NNX07AH11G and NSF AST-0707725. H.-J.
Seo is supported by the U.S. Department of Energy under contract
No. DE-AC02-07CH11359.



\clearpage

\newcommand{\tableskip}{\\[-8pt]}
\newcommand{\singleline}{\tableskip\hline\tableskip}
\newcommand{\doubleline}{\tableskip\hline\tableskip}
\newlength{\tablespread}\setlength{\tablespread}{30pt}
\newcommand{\dje}{\hspace{\tablespread}}
\tabletypesize{\small}
\def\arraystretch{1.1}

\begin{deluxetable}{c@{\dje}c@{\dje}cccc@{\dje}c@{\dje}c@{\dje}}
\tablewidth{405pt}
\tablecaption{\label{tab:hod_p} HOD properties}

\startdata \doubleline
Model   &Total \#   &Satellite  &$M_{cen}$	&$M_{sat}$	&$\bar{n}$ \footnotemark[2]      &$\bar{n}P_{0.2}$ &$\aleph_{nonlin}$ \\
    &of Galaxies\footnotemark[1]    &Fraction (\%)  &$(h^{-1}M_\odot)$ &$(h^{-1}M_\odot)$& ($h^{3} \Mpc^{-3}$)   & \\\singleline
DM &$4 \times 10^6$ &-&-&-&0.004	    &4.60   &0.0\\
HOD1  &$2 \times 10^6$    &5  &$1.4\times 10^{12}$&$9.2 \times 10^{13}$&0.002      &5.78   &450.0\\
HOD2  &$1 \times 10^6$    &5  &$2.6\times 10^{12}$&$1.5 \times 10^{14}$&0.001      &3.59   &700.0\\
HOD3  &$3 \times 10^5$    &5  &$6.4 \times 10^{12}$&$3.1 \times 10^{14}$&0.0003     &1.59   &1550.0
\enddata

\tablecomments{HODs are referred to by the designations under the
``Model'' heading throughout the paper.
\footnotetext[1]{Number of DM particles in the DM only case.}
\footnotetext[2]{The nominal Poisson shot-noise is $\bar{n}^{-1}$.}}
\end{deluxetable}

\begin{deluxetable}{c@{\dje}c@{\dje}c@{\dje}cc@{\dje}c@{\dje}cc}
\tablewidth{355pt}
\tablecaption{\label{tab:fit} Fit results for each HOD model}

\startdata \doubleline
Model   &$\alpha-1$     &$\sigma_\alpha$&$\Sigma_{nl}$& $r_{scale}$& $\chi^2$   &bias \footnotemark[1]   &$\sigma_8$ \\
    &(\%)       &(\%)       &&&(per d.o.f)    &(b)    &       \\\singleline
DM &0.0457     &0.2333	   &6.66   &19.99  &0.92       &1.25   &0.63       \\
HOD1  &0.1065     &0.2243     &5.61   &19.78  &0.94       &2.04   &1.11       \\
HOD2  &0.1634     &0.2449     &5.85   &19.98  &0.86       &2.28   &1.25       \\
HOD3  &0.4897     &0.3326     &6.27   &20.04  &0.72       &2.77   &1.55       
\tableskip
\enddata

\tablecomments{Fitting range: $30 \leq r_s \leq 200 h^{-1}\Mpc$.
$\sigma_\alpha$ is the error on the mean $\alpha$ of the 44 simulations.
\footnotetext[1]{Bias is not equal to $1$ for the DM only case because we are 
working in redshift space.}}
\end{deluxetable}

\clearpage
\begin{deluxetable}{c@{\dje}c}
\tablewidth{150pt}
\tablecaption{\label{tab:aldiff} Difference in mean $\alpha$ between $\wnot$
and $P(k)$}

\startdata \doubleline
Model   &$\langle \Delta\alpha_{\omega P} \rangle$     \\
    &(\%)               \\\singleline
DM &$0.0516 \pm 0.1205$	\\
HOD1  &$0.0076 \pm 0.0672$        \\
HOD2  &$0.0205 \pm 0.0600$        \\
HOD3  &$0.1035 \pm 0.0665$        
\enddata

\end{deluxetable}

\appendix

\section{Evaluating $\widetilde{W}_\ell(k)$} \label{sec:jayn}

The expressions for $\widetilde{W}_\ell(k)$, in terms of polynomials
of $k$ times trigonometric functions, involve a lot of cancellation.
This makes them unstable to direct evaluation.  However if we define
\begin{equation}
  K_n(k) = \frac{2+n}{(kr_s)^{2+n}} \int_0^{kr_s} x^n\sin x\ dx
\end{equation}
then
\begin{equation}
  \widetilde{W}_0(k) = \frac{8\pi}{3}
  \left[\frac{1}{3}\left(K_7-K_{16}\right)-\left(K_{10}-K_{13}\right)\right]
\end{equation}
while
\begin{equation}
  \widetilde{W}_2(k) = -\frac{24\pi}{5k^2}
  \left[ 3K_5-16K_8+25K_{11}-12K_{14} \right] .
\end{equation}
It is straightforward to evaluate $K_n(x)$, the limits are
\begin{equation}
  K_n(x) = 1 - \frac{n+2}{3!(n+4)}x^2 + \frac{n+2}{5!(n+6)}x^4 + \cdots
\end{equation}
as $x\to 0$ and
\begin{equation}
  K_n(x) = -(n+2)\frac{\cos x}{x^2} + n(n+2)\frac{\sin x}{x^3} + \cdots
\end{equation}
as $x\to \infty$.
The $K_n$ also satisfy a simple recurrence relation
\begin{equation}
  K_n(x) = \frac{n+2}{x^3}\left[ n\sin x-x\cos x-(n-1)xK_{n-2} \right] .
\end{equation}
Use of this recurrence relation for high $k$ and the power-series
expansion for low $k$ results in stable evaluation of the $\widetilde{W}_\ell$.

\end{document}